\RequirePackage{fix-cm}
\documentclass[smallextended]{svjour3}       
\smartqed  
\usepackage{mathptmx}  
\usepackage{cite}
\usepackage{graphicx}
\usepackage[numbers,sort&compress]{natbib}
\usepackage{amsmath}
\usepackage{amssymb}
\usepackage{amsmath}
\usepackage[utf8]{inputenc}
\usepackage{epstopdf}
\usepackage{natbib}
\usepackage{graphicx}
\usepackage[usenames,dvipsnames]{color}
\usepackage[usenames,dvipsnames,svgnames,table]{xcolor}
\definecolor{Violet}{RGB}{12,127,172}
\usepackage[colorlinks=true,linkcolor=Red,citecolor=Violet,urlcolor=Magenta]{hyperref}
\usepackage{color}
\hypersetup{
  colorlinks=true,
  citecolor=Violet,
  linkcolor=Red,
  urlcolor=Magenta}
\usepackage[numbers,sort&compress]{natbib}
\usepackage{mathtools}

\newcommand{\ead}{URL: }
\journalname{Journal on Wireless Personal Communications}
\begin{document}

\title{Achievable Rate Analysis of Cooperative Relay Assisted Opportunistic-NOMA under Rician Fading Channels with Channel State Information}

\author{Pranav Kumar Jha \and S Sushmitha Shree \and D. Sriram Kumar 
}
\institute{Pranav Kumar Jha \at
              Department of Electronics and Communication Engineering\\
              National Institute of Technology, Tiruchirappalli \\
              Tel.: +91-846-898-3890\\
              \email{jha\_k.pranav@live.com}  \\
         \ead{orcid.org/0000-0001-8053-988X}\\\\
         S Sushmitha Shree \at
              Department of Electronics and Communication Engineering\\
              Thiagarajar College of Engineering, Madurai \\
              \email{sushmithasriram@gmail.com}  \\
         \ead{orcid.org/0000-0003-4929-7776}
}
\date{Received: date / Accepted: date}

\maketitle
\begin{abstract}
The average achievable rate of an Opportunistic Non-Orthogonal Multiple Access (O-NOMA) based Cooperative Relaying System (CRS) with Channel State Information (CSI) known at the transmitter end is analyzed under Rician fading channels. For opportunistic transmission of data signals, CSI is used for the source-to-relay and source-to-destination links, which helps the transmitter to select the best out of the cooperative NOMA transmission and the direct transmission of signals, instantaneously. The average achievable rate of the O-NOMA and conventional NOMA (C-NOMA) based CRSs is considered for the performance comparison and the exact and asymptotic analytical expressions of the achievable rates has been provided. Asymptotic results are verified through Monte Carlo simulations for various channel powers and power allocation coefficients. Numerical results verify that the derived analytical results are matched well with the Monte Carlo simulations and show that O-NOMA-based CRS achieves better rate performance than C-NOMA-based CRS with the increasing power allocation coefficients, transmit Signal-to-Noise Ratios (SNRs) and channel powers.

\end{abstract}

\section{Introduction}
\label{sec1}
Non-orthogonal multiple access (NOMA) is one of the promising techniques to improve the spectral efficiency of wireless multi-user communication systems \cite{ding2014performance}. For multiple access, NOMA uses the power domain and superposition coding is used to implement this which enables the transmitter to transmit data signals with different power levels to different receivers at the same time and bandwidth \cite{do2009linear}. In order to boost the spectral efficiency of cooperative relaying systems (CRSs), the NOMA has been proposed into CRSs where one or more relays help a source in transmitting the signals to one or more destinations in order to mitigate the wireless channel effects such as path losses, attenuation and shadowing. In industry, the NOMA and the cooperative relaying technique have been considered as primary techniques for 3rd Generation Partnership Project (3GPP) Long Term Evolution-Advanced (LTE-A) systems \cite{danae1999technical,bai2013evolved}. Hence, CRSs using the NOMA can be practically realized as one of alternatives to achieve high spectral efficiency in the fifth generation (5G) communication systems. 

In literature, the advanced NOMA for different CRSs have been developed to significantly improve the spectral efficiency of the CRSs \cite{kim2015capacity,xu2016novel,ding2015cooperative,men2015performance,kim2015non,zhong2016non,luo2017adaptive,do2017bnbf,liu2016hybrid}. The research is focused on the models of CRSs with different node constructions and different relaying schemes and has shown the spectral efficiency gains achieved by their proposed NOMA schemes in terms of average achievable rate and outage probability. In \cite{kim2015capacity}, CRS using NOMA has been proposed, where the source
simultaneously transmits two independent data symbols by superposition coding, and the relay decodes and forwards the symbol with lower received power after performing the successive interference cancellation (SIC). Also, the average achievable rate for the CRS using NOMA has been analyzed, and a
sub-optimal power allocation scheme for NOMA has been presented. In \cite{xu2016novel}, a novel detection scheme for CRS using NOMA has been proposed to enhance the achievable rate for CRS using NOMA presented in \cite{kim2015capacity}, but it requires more complexity at the receiver. In \cite{xu2016novel}, the destination jointly decodes two data symbols from the direct and relayed transmissions with maximal-ratio combining and SIC. In
addition, its average achievable rate and the outage probability have been investigated. Unlike \cite{kim2015capacity} and
\cite{xu2016novel} considering a single relay and a single destination, in \cite{kim2015capacity}, the more complicated CRS with multiple
relays and destinations has been treated, and a cooperative NOMA scheme has been proposed for the CRS where the receivers with better channel conditions have prior information about the data symbols of other receivers and the prior information is used to achieve the spatial diversity. In \cite{ding2015cooperative}, the outage probability and diversity order achieved by the cooperative NOMA have been investigated. Unlike \cite{kim2015capacity,xu2016novel,ding2015cooperative} with decode-and-forward relaying, in \cite{men2015performance}, a NOMA-based
down-link cooperative cellular system with amplify-and-forward relaying has been proposed, where the base station transmits data signals to two users simultaneously with amplify-and-forward relaying. Also, its average achievable rate and outage performance have been investigated. In \cite{kim2015non}, the NOMA scheme has been proposed in coordinated direct and relay transmission (CDRT), where the base station directly communicates with a near user while communicating with a far user only through a relay. In the CDRT using NOMA, the near user has the prior information about the data symbol of the far user,
and exploits it for interference cancellation, which can significantly improve the spectral efficiency. Its
average achievable rate and outage probability have been also analyzed in \cite{kim2015non}. In \cite{zhong2016non}, the full-duplex
(FD) CRS using NOMA with dual users has been proposed and its average achievable rate and outage
probability have been investigated under the assumption of imperfect self-interference cancellation. In
addition, it has been shown that the FD CRS using NOMA can work better than the half-duplex CRS using NOMA. In the studies on CRSs using NOMA in \cite{kim2015capacity,xu2016novel,ding2015cooperative,men2015performance,kim2015non,zhong2016non}, channel state information (CSI) has been
assumed to be unavailable at the source in order to reduce the system overhead. However, it can limit the performance improvement because the source cannot use the time-varying channels for data transmission. In \cite{luo2017adaptive,do2017bnbf,liu2016hybrid}, hence, adaptive transmission, user selection, and hybrid relaying schemes using CSI have been respectively proposed to enhance the spectral efficiency of CRSs using NOMA. In \cite{lee2017achievable}, an opportunistic NOMA (O-NOMA) based
CRS has been proposed using decode-and-forward over Rayleigh fading channels and the average achievable rates has been studied for different channel powers and power allocation coefficients with CSI available at the transmitter end.

In this paper, unlike the proposed schemes in \cite{luo2017adaptive,do2017bnbf,liu2016hybrid,lee2017achievable}, we analyze the O-NOMA-based
CRS using CSI available at the transmitter end under Rician fading channels in order to achieve the further performance improvement at the expense of system overhead. We consider a source, a decode-and-forward relay, and a destination for the simple implementation as in \cite{kim2015capacity}. For opportunistic transmission, CSI for the source-to-destination and source-to-relay links is exploited and based
on the CSI, transmitter instantaneously selects one of the direct
transmission and the cooperative NOMA transmission \cite{kim2015capacity} with the help of the relay, which can provide better achievable rate performance than the conventional NOMA-based CRS with no CSI at the source. In addition, an asymptotic expression of the average
achievable rate has also been provided and the asymptotic results are verified through Monte Carlo simulations. The average achievable rates of the O-NOMA and the conventional NOMA (C-NOMA) based CRSs has been compared for different channel powers and power allocation coefficients used for NOMA. 

This paper is organized as follows: Section \ref{sec2} provides opportunistic NOMA-based CRS, and provides
its received signals and signal-to-noise ratios (SNRs). In Section {\ref{sec3}}, the average achievable rate for the O-NOMA-based CRS is analyzed and in section \ref{sec4}, an approximation method is used for asymptotic average achievable rate expressions. In Section \ref{sec5}, a comparison has been made between the O-NOMA and C-NOMA-based CRSs in terms of average achievable rate in order to verify the superiority of O-NOMA-based CRS over C-NOMA-based CRS through numerical results and simulations. Section \ref{sec6} concludes this paper. 

\begin{figure}
\includegraphics[width=0.75\textwidth]{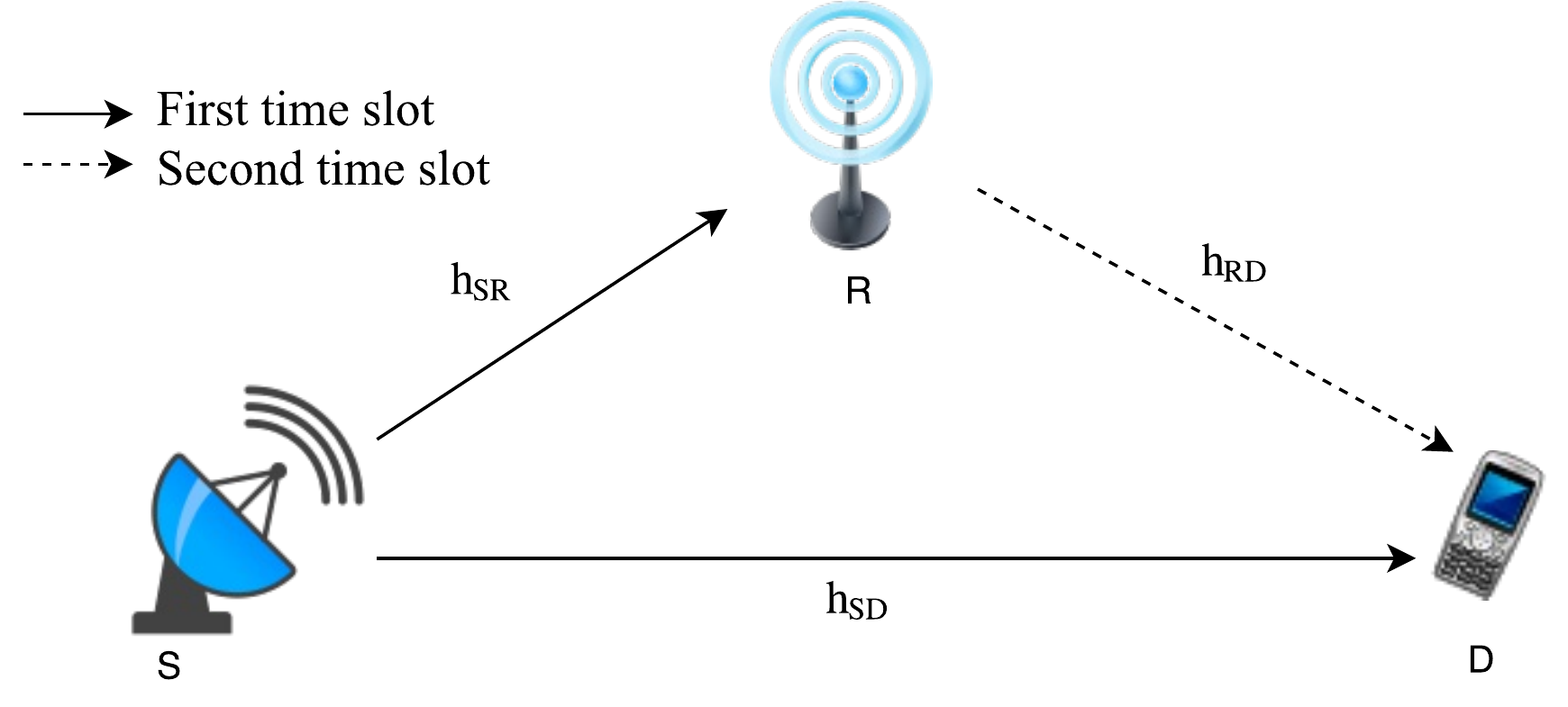}
\caption{A Cooperative relaying system based on NOMA.}
\label{fig1}
\end{figure}
\begin{figure}
\includegraphics[width=0.75\textwidth]{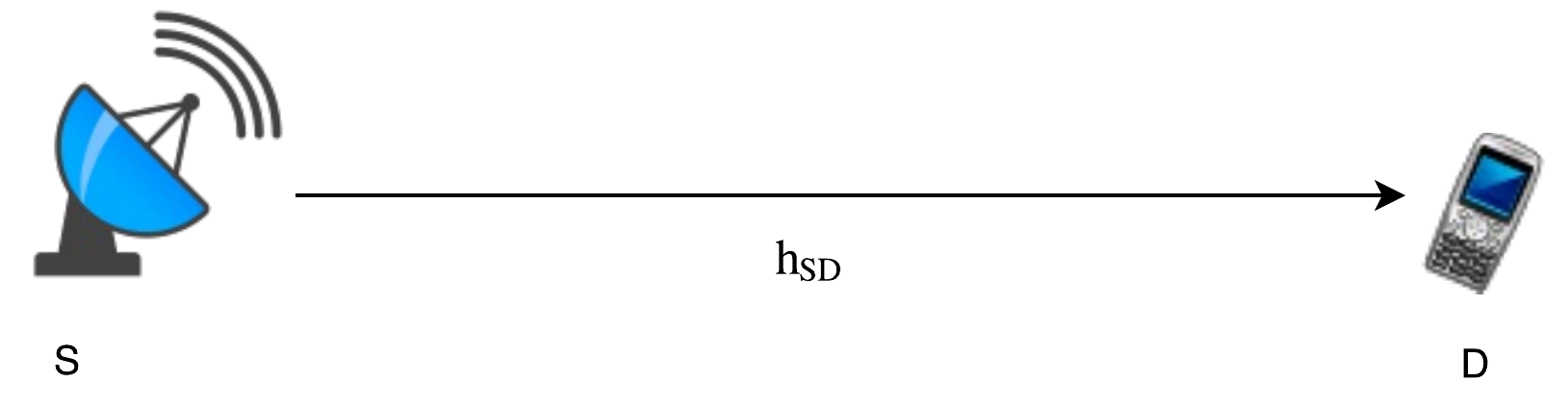}
\caption{Direct signal transmission from source to destination.}
\label{fig2}
\end{figure}
\section{System Description}
\label{sec2}
A CRS has been considered for the analysis as shown in Fig. \ref{fig1}, where a source (S) transmits signals to a destination (D)
directly and through a relay (R). The channel coefficients of S-to-D, S-to-R, and R-to-D links are represented as $h_{SD}$, $h_{SR}$, and $h_{RD}$, respectively, and considered to be independent Rician random variables
with average channel powers of $\Omega^2_{SD}$, $\Omega^2_{SR}$ and $\Omega^2_{RD}$, respectively.

\begin{figure}
\includegraphics[width=0.75\textwidth]{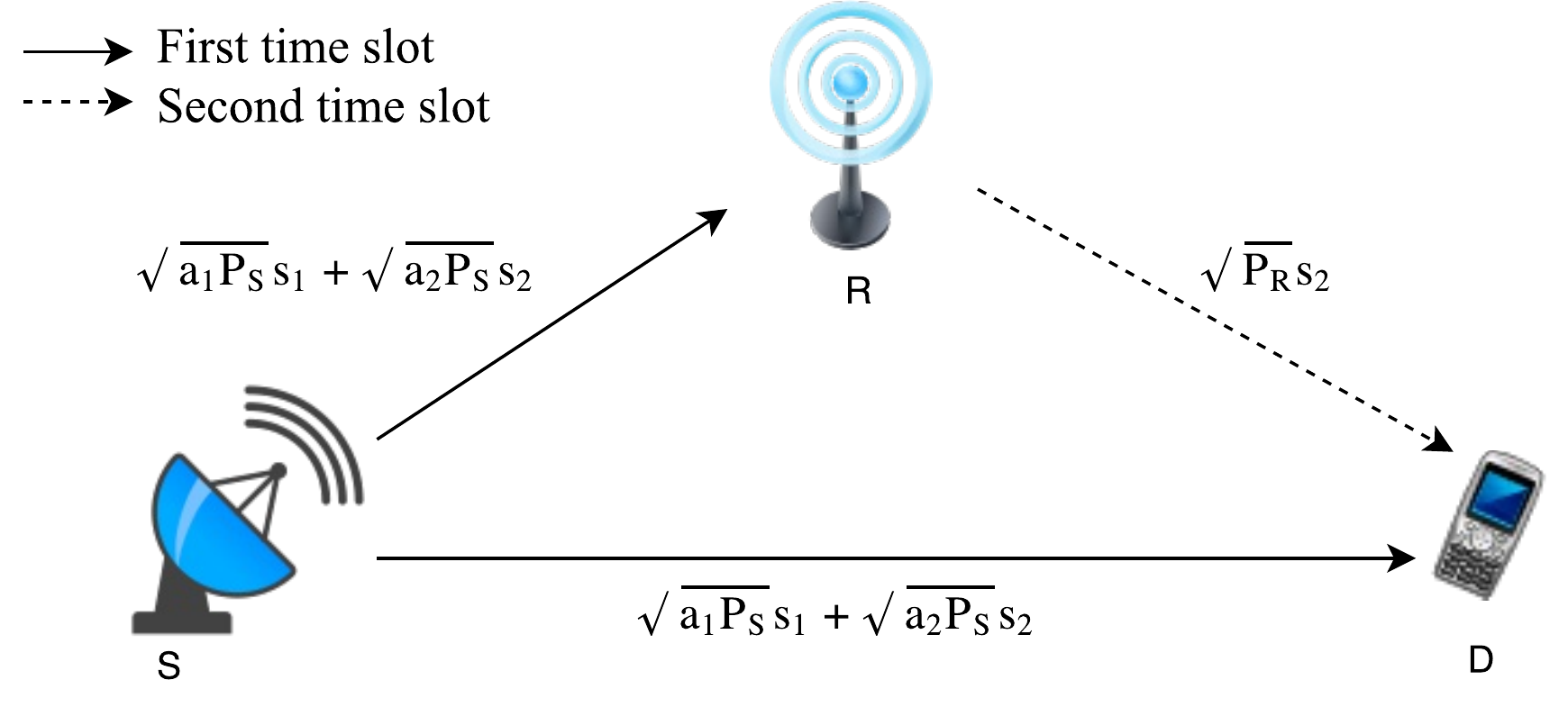}
\caption{NOMA-based cooperative relaying systems.}\label{fig3}
\end{figure}

In the CRS using NOMA, During the first time slot, the source transmits $\sqrt{a_1 P_S}s_1+\sqrt{a_2 P_S}s_2$ to the relay and the destination, using the superposition coding \cite{ding2014performance,kim2015capacity,xu2016novel} shown in Fig. \ref{fig3}, where $s_i$ represents the $i$-th data symbol, $E[{|s_i|}^2]=1$ and $P_S$ signifies the total transmit power, $a_i$ represents the power allocation coefficient for symbol $s_i$ and $a_1+a_2=1$ and $a_1 > a_2$ is assumed, which means
$0.5 < a_1 < 1$ and $0 < a_2 < 0.5$. The relay decodes symbol $s_2$ after decoding and cancelling symbol $s_1$ with SIC, whereas the destination decodes symbol $s_1$ considering symbol $s_2$ as noise. During the second
time slot, only the relay transmits the decoded symbol $s_2$ with power $P_S$ to the destination.
However, when $|h_{SD}|^2>|h_{SR}|^2$, the NOMA  may not provide a gain of end-to-end achievable rate since the received signal power of symbol  at the relay is very limited. For O-NOMA-based CRS, if $|h_{SD}|^2>|h_{SR}|^2$, only direct
transmission between the source and the destination is performed without relaying as shown in Fig. \ref{fig2} , otherwise the NOMA is employed with relay \cite{lee2017achievable}. The rationale of the O-NOMA-based CRS
is based on the fact that the achievable rate for the relayed link is the minimum of achievable
rates for the source-to-relay link and the relay-to-destination link. That is, when $|h_{SD}|^2>|h_{SR}|^2$, the
direct link achieves better rate performance than the relayed link. In the O-NOMA-based CRS, hence, the source
directly transmits symbol $s_1$ to the destination with power $P_S$ during a time slot when $|h_{SD}|^2>|h_{SR}|^2$.

In the C-NOMA-based CRS, the received signals at the relay and the destination during the first time
slot are respectively given as 
\begin{equation}
  r_{SR}^C  = h_{SR}(\sqrt{a_1 P_S}s_1+\sqrt{a_2 P_S}s_2) + n_{SR},
  \label{eq1}
\end{equation}
and
\begin{equation}
  r_{SD}^C=h_{SD}(\sqrt{a_1 P_S}s_1+\sqrt{a_2 P_S}s_2) + n_{SD},
   \label{eq2}
\end{equation}
where $n_{SR}$ and $n_{SD}$ denote additive white Gaussian noise with variance $\sigma^2$. The received SNRs of
symbol $s_1$ to be decoded for SIC and symbol $s_2$ to be decoded after SIC at the relay are respectively
given from (\ref{eq1}) as
\begin{equation}
  \gamma_{SR,s_1}^C=\frac{a_1 P_S|h_{SR}|^2}{a_2 P_S|h_{SR}|^2+\sigma^2}
   \label{eq3},
\end{equation}
and
\begin{equation}
  \gamma_{SR,s_2}^C=\frac{a_2 P_S|h_{SR}|^2}{\sigma^2}.
   \label{eq4}
\end{equation}

The received SNR of symbol $s_1$ to be decoded at the destination is given from (\ref{eq2}) as 
\begin{equation}
  \gamma_{SD, s_1}^C=\frac{a_1 P_S|h_{SD}|^2}{a_2 P_S|h_{SD}|^2+\sigma^2}.
   \label{eq5}
\end{equation}

The received signal at the destination during the second time slot is given as 
\begin{equation}
  r_{RD}^C=\sqrt{a_2 P_R}h_{RD}s_2+n_{RD},
     \label{eq6}
\end{equation}
where $n_{RD}$ is additive white Gaussian noise with variance $\sigma^2$, and thus the received SNR for symbol $s_2$ is given as 
\begin{equation}
  \gamma_{RD,s_2}^C=\frac{a_2 P_S|h_{RD}|^2}{\sigma^2}.
     \label{eq7}
\end{equation}

On the other hand, for the direct transmission, the received signal of symbol $s_1$ at the destination and its received SNR are respectively given as,
\begin{equation}
  r_{SD}^D=\sqrt{a_1 P_R}h_{SD}s_1+n_{SD},
  \label{eq8}
\end{equation}
\begin{equation}
  \gamma_{SD,s_1}^D=\frac{a_1 P_S|h_{SD}|^2}{\sigma^2}.
  \label{eq9}
\end{equation}
\section{Achievable Rate Analysis }
\label{sec3}
Let $\lambda_{SD}\triangleq|h_{SD}|^2$, $\lambda_{SR}\triangleq|h_{SR}|^2$, $\lambda_{RD}\triangleq|h_{RD}|^2$, $\rho=\frac{P_S}{\sigma^2}$ and $C(x)\triangleq log_2(1+x)$, where $\rho$
signifies the transmit SNR. It is assumed that \cite{kim2015capacity} the transmit power of S and R is the same as $P_S=P_R=P$. As following the fact the the end-to-end achievable rate of decode-and-forward relaying is dominated by the weakest link \cite{danae1999technical}, and with the help of \cite{danae1999technical,bai2013evolved,kim2015capacity,ding2015cooperative,kim2015non}, the achievable rate of the O-NOMA-based CRS is obtained as follows \cite{kim2015capacity}:
If $\lambda_{SD}<\lambda_{SR}$
\begin{equation}
\begin{split}
C^{Pro}=\frac{1}{2}\textrm{min}\{\textrm{log}_2(1+\gamma_{SD, s_1}^C),\textrm{log}_2(1+\gamma_{SR,s_1}^C)\}\\ + \frac{1}{2}\textrm{min}\{\textrm{log}_2(1+\gamma_{SR, s_2}^C),\textrm{log}_2(1+\gamma_{RD,s_2}^C)\},
\end{split}
  \label{eq10}
\end{equation}
which can be written as \cite{lee2017achievable}
\begin{equation}
\begin{split}
C^{Pro} & =\frac{1}{2}\textrm{min}\{\textrm{log}_2(1+\gamma_{SD, s_1}^C)\}\\
& + \frac{1}{2}\textrm{min}\{\textrm{log}_2(1+\gamma_{SR, s_2}^C),\textrm{log}_2(1+\gamma_{RD,s_2}^C)\},
\end{split}
  \label{eq11}
\end{equation}
else,
\begin{equation}
C^{Pro}= \textrm{log}_2(1+\gamma_{SD,s_1}^D).
  \label{eq12}
\end{equation}

In (\ref{eq10}), it is noted that the first and the second parts denote the achievable rates of symbols  $s_1$ and $s_2$, respectively. In the first part of (\ref{eq10}), $log_2(1+\gamma_{SR,s_1}^C)$ is required to assume that the relay successfully decodes symbol  $s_1$ for SIC, but that is deleted in (\ref{eq11}) since  $\lambda_{SD}<\lambda_{SR}$. It is also noted that there is one half spectral efficiency penalty for relaying in (\ref{eq11}), whereas there is no spectral efficiency penalty in (\ref{eq12}) as the source transmits an independent data symbol to the destination directly for a given time slot when  $\lambda_{SD}>\lambda_{SR}$. Then, using (\ref{eq11}) and (\ref{eq12}), the achievable average rate of the O-NOMA-based CRS is obtained as
\begin{equation}
\begin{split}
C^{Pro} & = \underbrace{\frac{1}{2}\textrm{log}_2(1+{\lambda_{SD}\rho)}-\frac{1}{2}(1+a_2 \lambda_{SD} \rho)}_{C_{C,{s_1}}}\\
& + \underbrace{\frac{1}{2}\textrm{log}_2(1+\textrm{min}\{a_2\lambda_{SR},\lambda_{RD}\} \rho)}_{C_{C,{s_2}}},
\end{split}  \label{eq13}
\end{equation}
else,
\begin{equation}
C^{pro}= \underbrace{\textrm{log}_2(1+\lambda_{SD}\rho)}_{C_{D,{s_1}}}.
  \label{eq14}
\end{equation}

With the help of (\ref{eq13}) and (\ref{eq14}), the total achievable rate of O-NOMA-based CRS can be written as 
\begin{equation}
\begin{split}
C^{\overline{Pro}} & =\frac{1}{2}\textrm{log}_2(1+{\lambda_{SD}\rho)}-\frac{1}{2}(1+a_2 \lambda_{SD} \rho
)\\
& + \frac{1}{2}\textrm{log}_2(1+\textrm{min}\{a_2\lambda_{SR},\lambda_{RD}\} \rho)\\
& + \textrm{log}_2(1+\lambda_{SD}\rho).
\end{split}  \label{eq15}
\end{equation}

Let $\gamma_1\triangleq \lambda_{SD}$,  $\gamma_2\triangleq \textrm{min}{\{\lambda_{SR},\lambda_{RD}\}}$, the cumulative distribution function (CDF) of $\gamma_1$ and $\gamma_2$ for $C_{C,{s_1}}$ and $C_{C,{s_2}}$ is given as \cite{jiao2017performance}
\begin{equation}
\begin{split}
F(\gamma_1) & = 1-A_x A_y\sum_{k=0}^{\infty}\sum_{n=0}^{\infty}\tilde B_x(n)\tilde B_y(k) n!k!e^{-(a_x+a_y)\gamma_1}\sum_{i=0}^{n}\sum_{k=0}^{k} \frac{a_x^ia_y^j}{i!j!}\gamma_1^{i+j},
\end{split}\label{eq17}
\end{equation}
\begin{equation}\begin{split}
F(\gamma_2) & = 1-A_z A_y\sum_{k=0}^{\infty}\sum_{n=0}^{\infty}\tilde B_z(n)\tilde B_y(k) n!k!e^{-(a_z+\frac{a_y}{a_2})\gamma_2} 
\sum_{i=0}^{n}\sum_{k=0}^{n} \frac{a_z^i {(a_y/a_2)}^j}{i!j!}\gamma_2^{i+j},
\end{split}\label{eq18}
\end{equation}
where $B_x(n)=\frac{K_x^n(1+K_x)^n}{\Omega_x^n(n!)^2}, B_y(k)=\frac{K_y^k(1+K_y)^k}{\Omega_y^k(k!)^2}, a_x=\frac{1+K_x}{\Omega_x},  a_y=\frac{1+K_y}{\Omega_y}, A_x=a_xe^{-K_x}, A_y=a_ye^{-K_y},$
$\tilde B_x(n)=\frac{B_x(n)}{a_x^{n+1}},
B_y(k)=\frac{B_y(k)}{a_y^{k+1}}$. Here, x represents the
S-to-D link, y represents the S-to-R link, z represents the R-to-D
link and $K$ represents the Rician factor. The expansion form
of incomplete Gamma function is used for the equality
of (\ref{eq17}).

\section{Achievable Rate Approximation}
\label{sec4}
According to \cite{jiao2017performance}, \eqref{eq17} and \eqref{eq18} can be approximated for the asymptotic results where approximated values for $C_{C,{s_1}}$ and $C_{C,{s_2}}$ is given as
\begin{equation}
C_{C,{s_1}}=\frac{1}{2\textrm{ln}(2)}[H(\rho)-H(\rho a_2)],\label{eq28}
\end{equation}
where,
\begin{equation}
\begin{split}
H(\rho) & = A_x A_y\sum_{k=0}^{\infty}\sum_{n=0}^{\infty}\tilde B_x(n)\tilde B_y(k) n!k!
\sum_{i=0}^{n}\sum_{j=0}^{k} \frac{(i+j)!}{i!j!}\frac{a_x^ia_y^j}{\rho^{i+j}}e^\frac{a_x+a_y}{\rho}\bigg(\frac{1}{2\frac{a_x+a_y}{\rho}}\bigg)^{i+j}\\ & \times
\frac{\pi}{n}\sum_{t=1}^{n}(\textrm{cos}(\frac{2t-1}{2n}\pi)+1)^{i+j-1}
e^{-\frac{2\frac{a_x+a_y}{\rho}}{\textrm{cos}(\frac{2t-1}{2n}\pi)+1}}|\textrm{sin}(\frac{2t-1}{2n}\pi)|,
\end{split}\label{eq29}
\end{equation}
Similarly $C_{D,{s_1}}$ can be calculated as
\begin{equation}
 C_{D,{s_1}}=\frac{1}{\textrm{ln}(2)}H(\rho),\label{eq30}   
\end{equation}
 and
\begin{equation}C_{C,{s_2}}=\frac{1}{2\textrm{ln}(2)}G(\rho),
\label{eq31}
\end{equation}
where, 
\begin{equation}
\begin{split}
G(\rho) & = A_z A_y\sum_{k=0}^{\infty}\sum_{n=0}^{\infty}\tilde B_z(n)\tilde B_y(k) n!k!\sum_{i=0}^{n}\sum_{j=0}^{k} \frac{(i+j)!}{i!j!}\frac{a_z^i{(a_y/a_2)}^j}{\rho^{i+j}}e^\frac{a_z+a_y/a_2}{\rho} \\
& \times \bigg(\frac{1}{2\frac{a_z+a_y/a_2}{\rho}}\bigg)^{i+j} \frac{\pi}{n}
\sum_{t=1}^{n}(\textrm{cos}(\frac{2t-1}{2n}\pi)+1)^{i+j-1}
e^{-\frac{2\frac{a_z+a_y/a_2}{\rho}}{\textrm{cos}(\frac{2t-1}{2n}\pi)+1}}|\textrm{sin}(\frac{2t-1}{2n}\pi)|.
\end{split}\label{eq32}
\end{equation}
\section{Numerical Results and Discussions}
\label{sec5}
\begin{figure}
\includegraphics[width=0.75\textwidth]{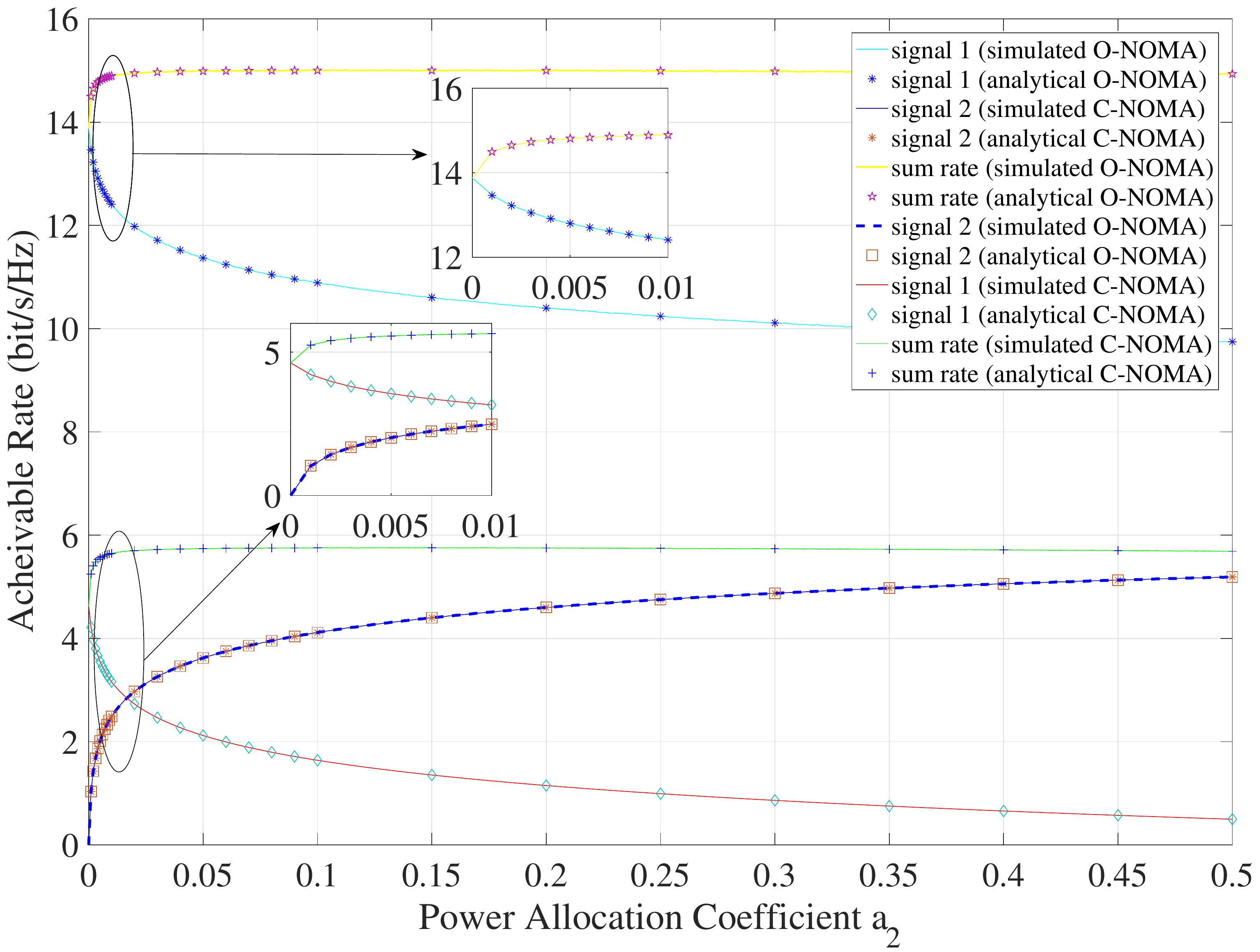}
\caption{Average achievable rates of O-NOMA and C-NOMA-based CRSs for $\rho=20$ dB when $\Omega_{SD}=3$ and $\Omega_{SR}=\Omega_{RD}=6$.}
\label{fig4}
\end{figure}
\begin{figure}
\includegraphics[width=0.75\textwidth]{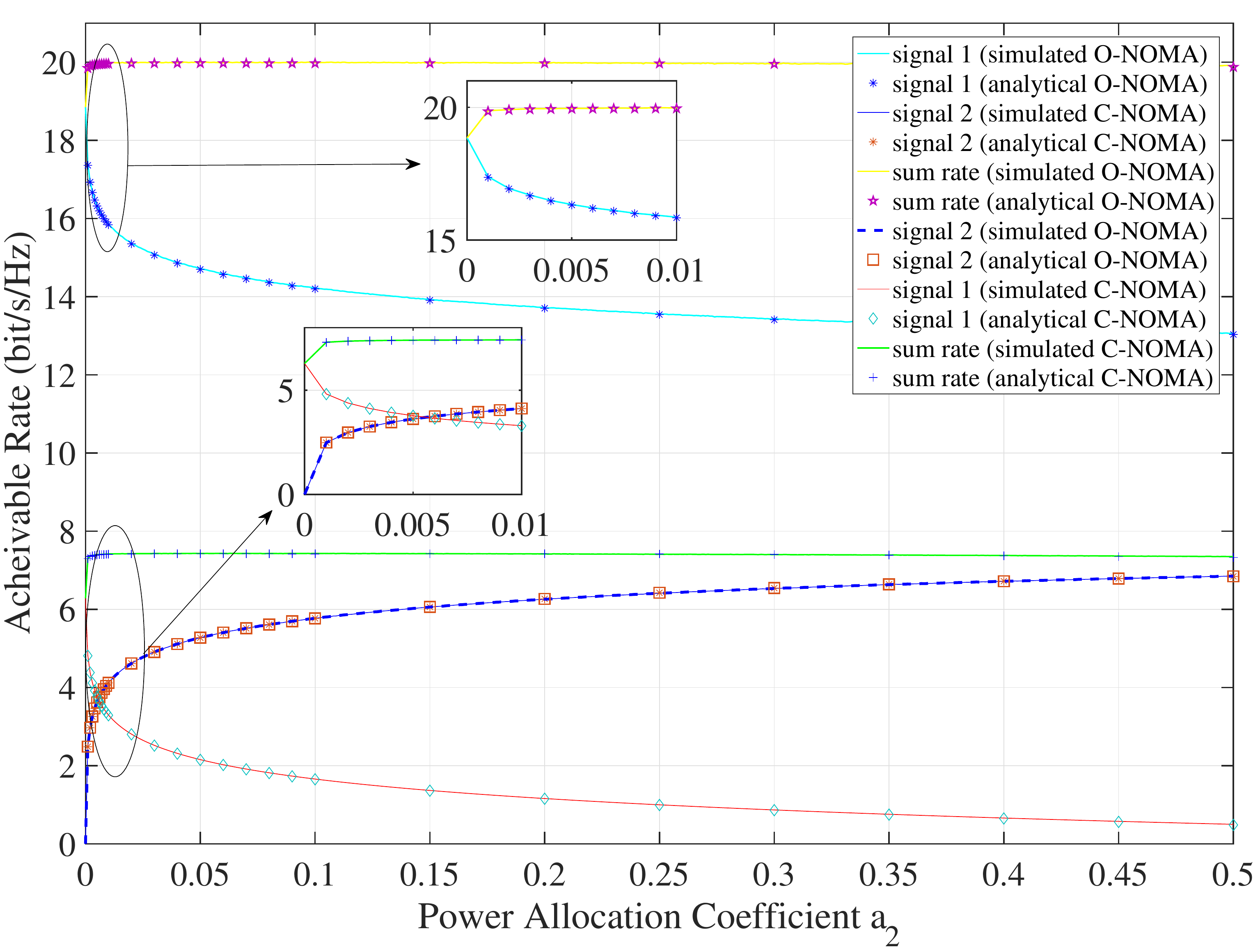}
\caption{Average achievable rates of O-NOMA and C-NOMA-based CRSs for $\rho=30$ dB when $\Omega_{SD}=3$ and $\Omega_{SR}=\Omega_{RD}=6$.}
\label{fig5}
\end{figure}
\begin{figure}
\includegraphics[width=0.75\textwidth]{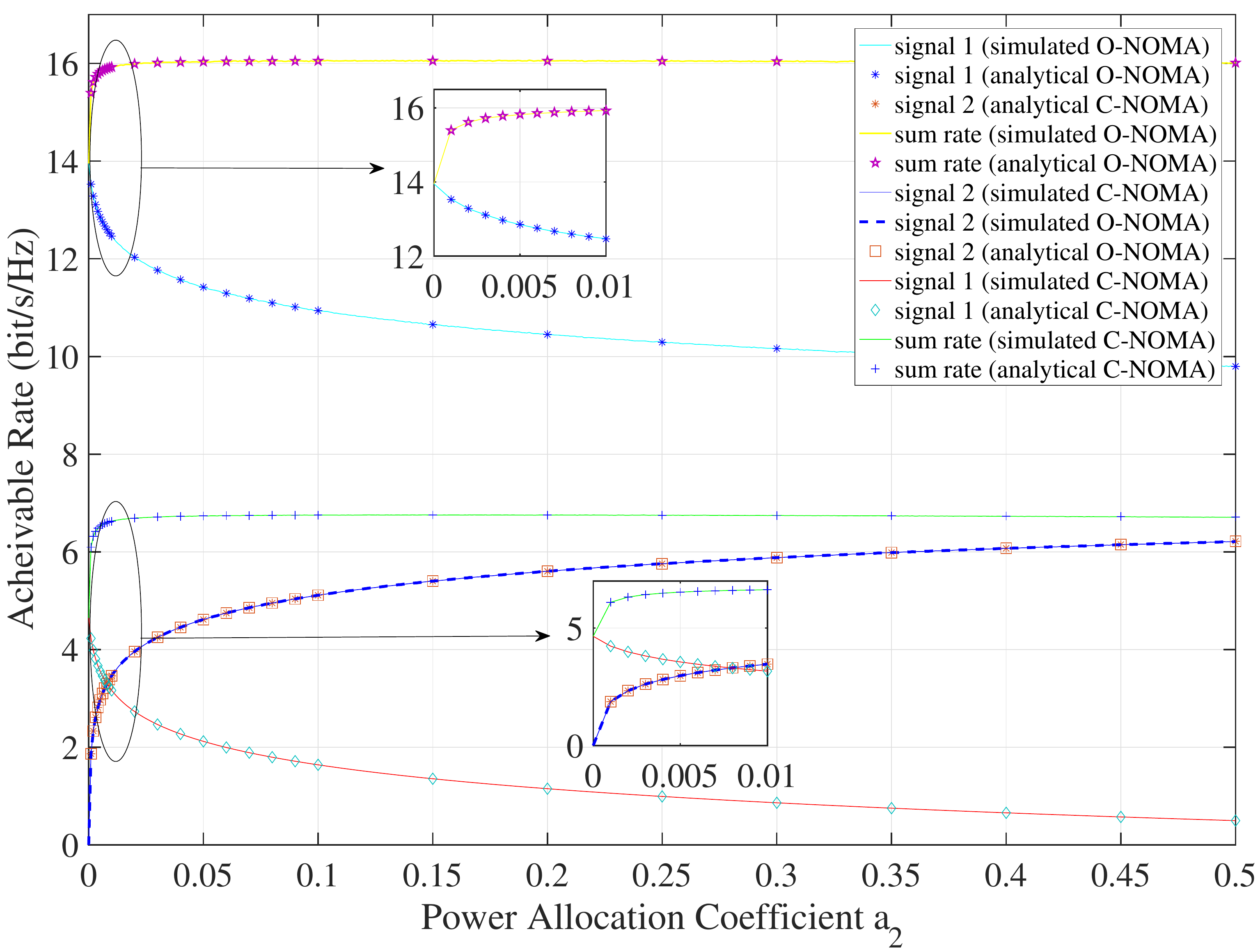}
\caption{Average achievable rates of O-NOMA and C-NOMA-based CRSs for $\rho=20$ dB when $\Omega_{SD}=3$ and $\Omega_{SR}=\Omega_{RD}=12$.}
\label{fig6}
\end{figure}
\begin{figure}
\includegraphics[width=0.75\textwidth]{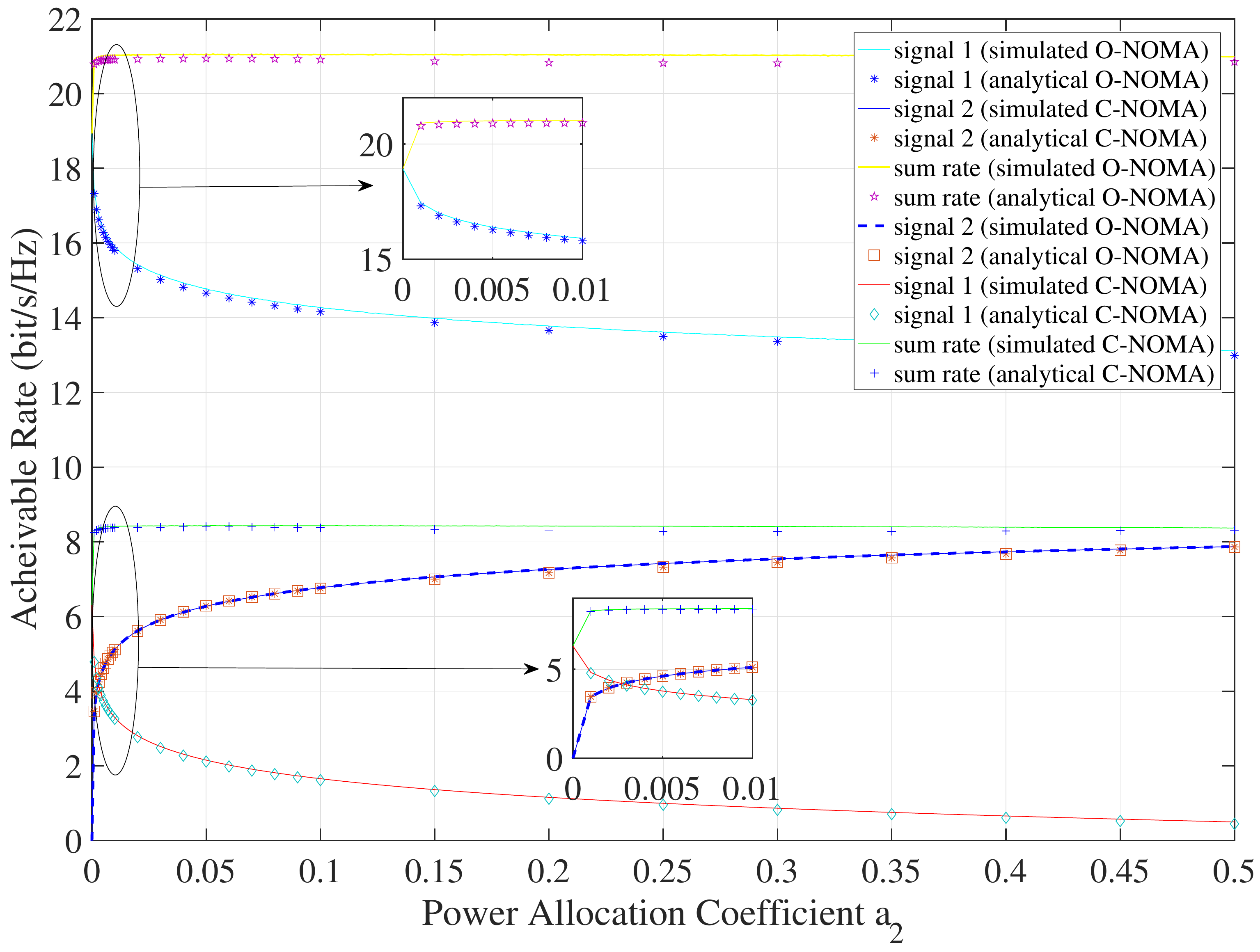}
\caption{Average achievable rates of O-NOMA and C-NOMA-based CRSs for $\rho=30$ dB when $\Omega_{SD}=3$ and $\Omega_{SR}=\Omega_{RD}=12$.}
\label{fig7}
\end{figure}
\begin{figure}
\includegraphics[width=0.75\textwidth]{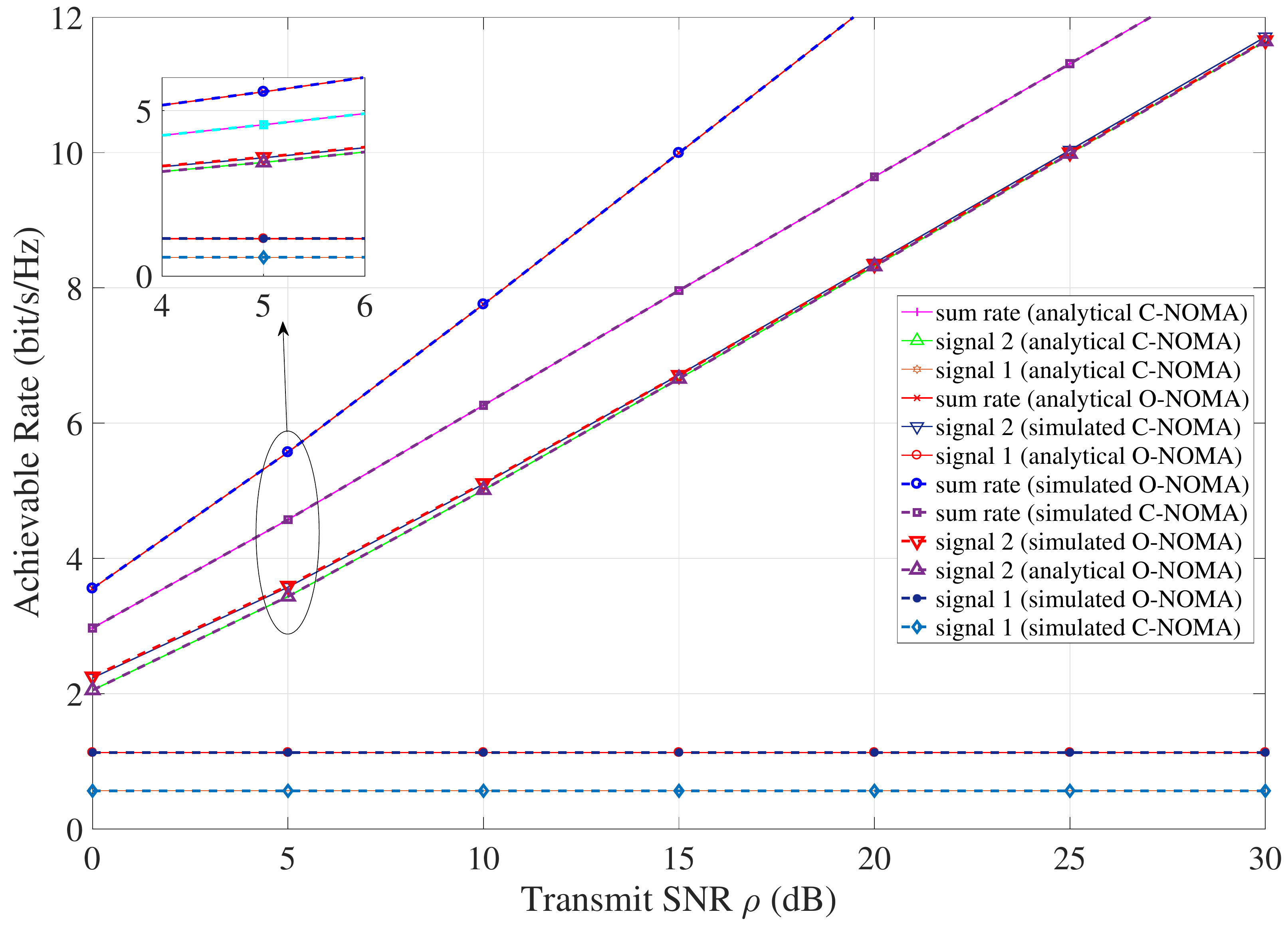}
\caption{Average achievable rates of O-NOMA and C-NOMA-based CRSs for $a_1=0.9$ and $a_2=0.1$ when $\Omega_{SD}=3$ and $\Omega_{SR}=\Omega_{RD}=6$.}
\label{fig8}
\end{figure}
\begin{figure}
\includegraphics[width=0.75\textwidth]{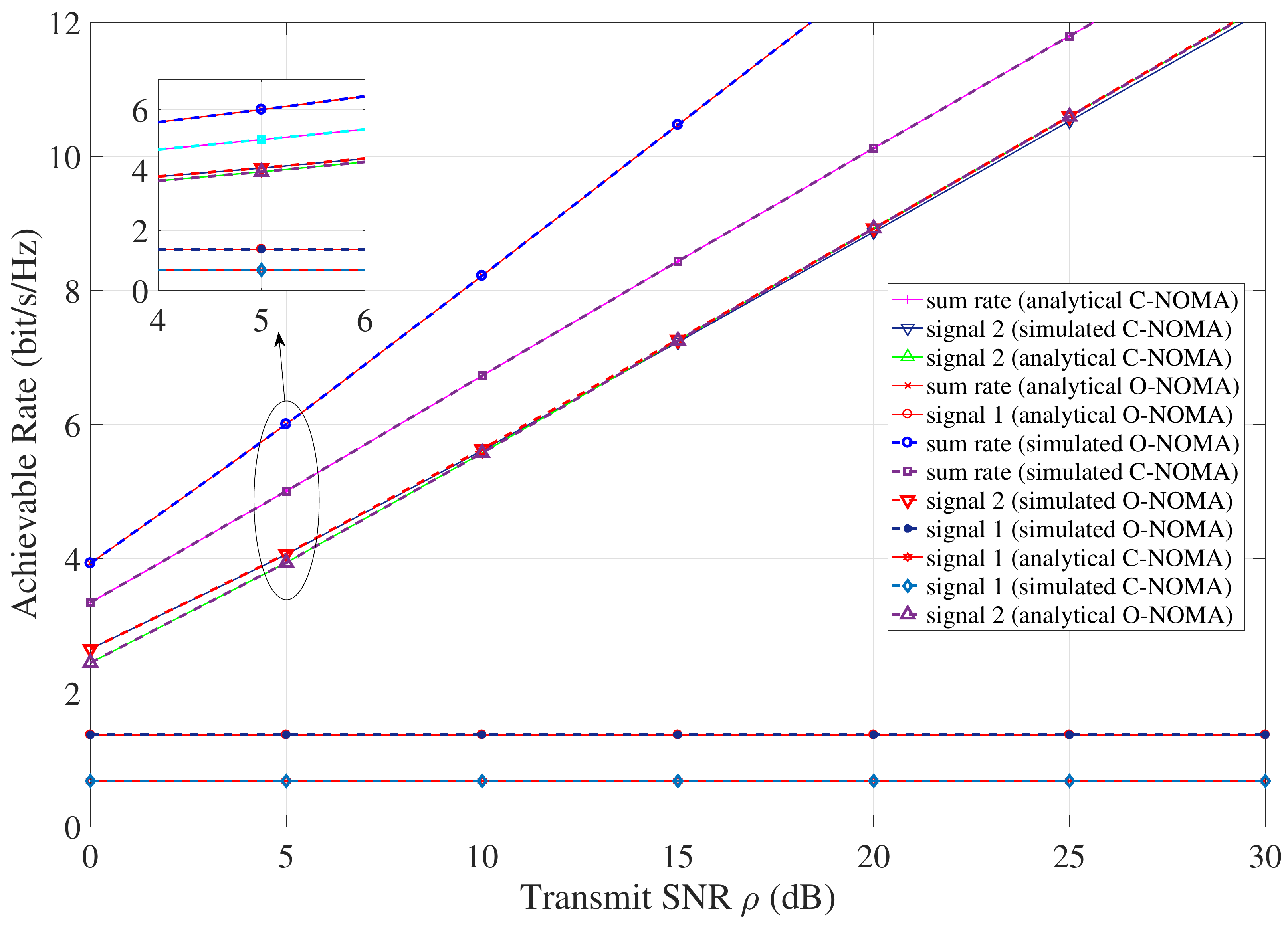}
\caption{Average achievable rates of O-NOMA and C-NOMA-based CRSs for $a_1=0.9$ and $a_2=0.1$ when $\Omega_{SD}=3$, $\Omega_{SR}=\Omega_{RD}=12$.}
\label{fig9}
\end{figure}

In this section, for validation of the accuracy of the analytical results analyzed in the Section \ref{sec4} has been compared using Monte Carlo simulations. For simulation purpose, $10^5$ realizations of Rician distribution random variables are transmitted. The asymptotic achievable rate of the C-NOMA-based CRS given in \cite{jiao2017performance} is used for comparison purpose with the achievable rate of O-NOMA-based CRS obtained in this paper.
Figs. \ref{fig4}, \ref{fig5}, \ref{fig6} and \ref{fig7} provides the achievable rate performance of $s_1$, $s_2$ and the corresponding average achievable sum rate of the O-NOMA and C-NOMA-based CRSs for different values of the power allocation coefficient $a_2$ of $s_2$. Further, parameters are set as $K_{SR} = K_{RD} = 5$, $K_{SD} = 2$ and $\Omega_{SD} = 3$. In addition, as $a_2$ increases, the achievable rate for $s_2$ increases and the rate for $s_1$ decreases, since $s_2$ gets more power.
Fig. \ref{fig8} and Fig. \ref{fig9} provides the achievable rate performance of the O-NOMA and C-NOMA-based CRSs against the transmit SNR $\rho$ where $a_2$ is set as 0.4.

In Fig. \ref{fig4}, for SNR = 20 dB and $\Omega_{SR} = \Omega_{RD} = 6$, It can be seen that when $a_2$ increases from 0.1 to 0.4, the achievable rate of $s_1$ decreases from 10.89 bit/s/Hz to 9.905 bit/s/Hz while for C-NOMA-based CRS, $s_1$ decreases from 1.64 bit/s/Hz to 0.657 bit/s/Hz. Next, $s_2$ increases from 4.113 bit/s/Hz to 5.058 bit/s/Hz for O-NOMA-based CRS which is same as in the case of C-NOMA-based CRS due to the available CSI at the transmitter end. Average achievable sum rates for O-NOMA and C-NOMA-based CRSs are 15 bit/s/Hz and 5.753 bit/s/Hz, respectively for $a_2=0.1$. 

In Fig. \ref{fig5}, SNR is set as 30 dB and $\Omega_{SR} = \Omega_{RD} = 6$.
Now when $a_2$ increases from 0.1 to 0.4, the achievable rate of $s_1$ decreases from 14.2 bit/s/Hz to 13.2 bit/s/Hz for O-NOMA-based CRS while for C-NOMA-based CRS, $s_1$ decreases from 1.651 bit/s/Hz to 0.6489 bit/s/Hz. Further, $s_2$ increases from 5.77 bit/s/Hz to 6.721 bit/s/Hz for O-NOMA and C-NOMA-based CRSs both. Average achievable sum rates for O-NOMA and C-NOMA-based CRSs are 19.97 bit/s/Hz and 7.421 bit/s/Hz, respectively for $a_2=0.1$.

In Figs. \ref{fig4} and \ref{fig5}, for $\Omega_{SR} = \Omega_{RD} = 6$, an increase of 4.97 bit/s/Hz for O-NOMA-based CRS and 1.668 bit/s/Hz for C-NOMA-based CRS has been shown in the average achievable sum rates for 10 dB increment in the SNR. O-NOMA-based CRS achieves 9.247 bit/s/Hz more rate than C-NOMA-based CRS for SNR=20 dB and 12.549 bit/s/Hz for SNR=30 dB at $a_2=0.1$, which shows that O-NOMA-based CRS achieves better average rate performance than C-NOMA-based CRS. 

In Fig. \ref{fig6}, SNR is set as 20 dB and $\Omega_{SR} = \Omega_{RD} = 12$.
Now, when $a_2$ increases from 0.1 to 0.4, the achievable rate of $s_1$ decreases from 10.94 bit/s/Hz to 9.957 bit/s/Hz for O-NOMA-based CRS while for C-NOMA-based CRS, $s_1$ decreases from 1.64 bit/s/Hz to 0.657 bit/s/Hz and $s_2$ increases from 5.112 bit/s/Hz to 6.074 bit/s/Hz for both the cases. Average achievable rates for O-NOMA and C-NOMA-based CRSs are 16.05 bit/s/Hz and 6.752 bit/s/Hz, respectively for $a_2=0.1$.

In Fig. \ref{fig7}
by using SNR = 30 dB and $\Omega_{SR} = \Omega_{RD} = 12$,
when $a_2$ increases from 0.1 to 0.4, the achievable rate of $s_1$ decreases from 14.16 bit/s/Hz to 13.15 bit/s/Hz for O-NOMA-based CRS while for C-NOMA-based CRS, $s_1$ decreases from 1.619 bit/s/Hz to 0.6115 bit/s/Hz and $s_2$ increases
from 6.756 bit/s/Hz to 7.678 bit/s/Hz for O-NOMA and C-NOMA-based CRSs both. Average achievable sum rates for O-NOMA and C-NOMA-based CRSs are 20.91 bit/s/Hz and 8.375 bit/s/Hz, respectively for $a_2=0.1$.

In Figs. \ref{fig6} and \ref{fig7}, for  $\Omega_{SR} = \Omega_{RD} = 12$, an increase of 4.86 bit/s/Hz for O-NOMA-based CRS and 1.619 bit/s/Hz for C-NOMA-based CRS has been shown in the average achievable sum rates for 10 dB increment in the SNR. O-NOMA-based CRS achieves 9.298 bit/s/Hz more rate than C-NOMA-based CRS for SNR=20 dB and 12.539 bit/s/Hz for SNR=30 dB at $a_2=0.1$ which shows that O-NOMA-based CRS achieves better average rate performance than C-NOMA-based CRS.

In Figs. \ref{fig4} and \ref{fig6}, for $a_2=0.1$ and SNR = 20 dB,  when values of $\Omega_{SR} = \Omega_{RD}$ changes from 6 to 12, an increase of 1.05 bit/s/Hz and 0.999 bit/s/Hz in the average achievable sum rate of O-NOMA-based CRS and C-NOMA-based CRS has been noticed, respectively and in Figs. \ref{fig5} and \ref{fig7}, for $a_2=0.1$ and SNR = 30 dB, when values of $\Omega_{SR} = \Omega_{RD}$ changes from 6 to 12, an increase of 0.94 bit/s/Hz and 0.954 bit/s/Hz in the average achievable sum rate of O-NOMA-based CRS and C-NOMA-based CRS has been shown, respectively. Hence. these results clearly demonstrate that larger channel power helps in achieving better achievable rate.

In Fig. \ref{fig8}, $a_1=0.9$, $a_2=0.1$ and $\Omega_{SR} = \Omega_{RD} = 6$ and when $\rho$ increases from 5 dB to 15 dB, the average achievable sum rate of O-NOMA-based CRS increases from 5.571 bit/s/Hz to 9.995 bit/s/Hz and for C-NOMA-based CRS, it increases from 4.575 bit/s/Hz to 7.961 bit/s/Hz while $s_2$ increases from 3.438 bit/s/Hz to 6.658 bit/s/Hz for O-NOMA and C-NOMA-based CRSs both due to the knowledge of CSI at the transmitter end. At $\rho=15$ dB, $s_1$ is 1.133 bit/s/Hz for O-NOMA-based CRS and 0.5663 bit/s/Hz for C-NOMA-based CRS. 

In Fig. \ref{fig9}, parameters are set as $\Omega_{SR} = \Omega_{RD} = 12$, $a_1=0.9$ and $a_2=0.1$.
Here, when $\rho$ increases from 5 dB to 15 dB, the average achievable sum rate of O-NOMA-based CRS increases from 6.007 bit/s/Hz to 10.47 bit/s/Hz and for C-NOMA-based CRS, increases from 5.011 bit/s/Hz to 8.44 bit/s/Hz while $s_2$ increases from 3.942 bit/s/Hz to 8.44 bit/s/Hz for O-NOMA and C-NOMA-based CRSs both. At $\rho=15$ dB, $s_1$ is 1.376 bit/s/Hz for O-NOMA-based CRS and 0.6879 bit/s/Hz for C-NOMA-based CRS.

In Figs. \ref{fig8} and \ref{fig9}, at $\rho=15$ dB, an increase of 0.475 bit/s/Hz for O-NOMA-based CRS and 0.479 bit/s/Hz for C-NOMA-based CRS has been shown in the average achievable sum rate  when $\Omega_{SR} = \Omega_{RD}$ varies from 6 to 12. O-NOMA-based CRS achieves 2.034 bit/s/Hz more rate than C-NOMA-based CRS for $\Omega_{SR} = \Omega_{RD}=6$ and 2.03 bit/s/Hz for $\Omega_{SR} = \Omega_{RD}=12$, which shows that O-NOMA-based CRS achieves better average rate performance than C-NOMA-based CRS and larger channel power helps in achieving improved achievable rate comparatively.
\section{Conclusions}
\label{sec6}
In this paper, the performance of O-NOMA-based CRS is investigated and exact and asymptotic analytical expressions of the achievable rates has been provided under Rician fading channels. Derived analytical results have been verified with the simulation results and proved to be in harmony with the Monte Carlo simulations which concludes that O-NOMA-based CRS is superior than C-NOMA-based CRS in terms of average achievable rate. Numerical results also show that as the transmit SNR increases and the channel power of the S-to-R and R-to D link is large, O-NOMA-based CRS achieves better rate performance than conventional C-NOMA-based CRS.
\bibliographystyle{spphys}       
\bibliography{references}
\end{document}